\documentclass[twocolumn,showpacs]{revtex4}
\usepackage{graphicx}
\usepackage{amsmath}
\usepackage{amssymb,amsthm}
\graphicspath{{pict/}{}}
\usepackage{bm}

\newcounter{Fig}

\newcommand{\be}{\begin{equation}}
\newcommand{\ee}{\end{equation}}

\begin{document}

\title{Embedded states in the continuum for ${\mathcal{P T}}$-symmetric systems}

\author{Mario I. Molina$^{1}$ and Yuri S. Kivshar$^{2}$}

\affiliation{$^{1}$Departamento de F\'{\i}sica, MSI-Nucleus on Advanced Optics,  and Center for Optics and Photonics (CEFOP), 
Facultad de Ciencias, Universidad de Chile, Santiago, Chile.\\
$^{2}$ Nonlinear Physics Centre, Research School of Physics and Engineering, 
Australian National University, Canberra ACT 0200, Australia}

\begin{abstract}
We introduce the novel concept of a bound state in the continuum (BIC) for 
a binary lattice satisfying the ${\mathcal{P T}}$ symmetry condition. We show how to 
build such state and the local potential necessary to sustain it. We find that an appropriate choice of the envelope function can bring the system from a ${\mathcal{P T}}$-symmetric phase into a Hamiltonian one. For more general envelope functions, the BIC can still be created but the bounded state 
will force the system to undergo the ${\mathcal{P T}}$ symmetry breaking transition.

\end{abstract}

\maketitle
 
\section{Introduction}
It is common knowledge that the general structure of a quantum system with a finite potential consists of bound states which are localized and normalizable, and unbound states which are extended and non-normalizable. The latter ones have positive energies if we consider the potentials vanishing at infinity.

In 1929 Wigner and von Neumann claimed an exception to this picture by building explicitly 
a bound state embedded in the continuum (BIC). It consisted of an eigenstate with positive energy but localized in space and square-integrable\cite{wigner}. What they did was to impose a modulation on a selected extended eigenstate, by means of a decaying envelope. With this,
the shape of  the local potential needed to generate such state was obtained. Both, the potential and the wavefunction thus proposed, oscillated in space and decayed as a power law.
The idea was forgotten for many years, until it was retaken again by Stillinger and coauthors\cite{pra10_1122,pra11_446} who suggested that BICs might be found in certain atomic and molecular systems. Later, they went on to suggest the use of superlattices to construct potentials that could support BICs\cite{stillinger_physB,herrick_physB}. Later, experiments with semiconductor heterostructures provided the direct observation of electronic bound states above a potential well localized by Bragg reflections\cite{capasso}. 

A different way to approach the problem of building potentials than can support BICs, come from the concept of resonant states in quantum mechanics. Resonant states are localized in space but with energies in the continuum, and they eventually decay, i.e., they have a finite lifetime. Under certain conditions, the interference between resonances can lead to a resonant state of zero width where the localized state decouples from the continuum becoming a BIC. One example of  this is the case of an Hydrogen atom in a uniform magnetic field, modeled as a system of coupled Coulombic channels, where interference between resonances belonging to different channels leads to the ocurrence of BICs\cite{coulombic}. More recently, BICs have been shown to occur in mesoscopic electron transport and quantum waveguides\cite{mesoscopic}, and in quantum dot systems\cite{qdot}. In this case, the existence of BICs can be traced back to the destruction of the discrete-continuum decay channels by quantum interference effects.

On the other hand, it has been admitted that the ultimate origin of the BIC phenomenon rests on   interference and thus, is inherent to any wave-like theory besides quantum mechanics, such as optical systems described by the paraxial wave equation. In fact, the analogy between these two fields have gained much attention recently, and have given rise to experimental observations of many phenomena that are hard to observe in a condensed matter setting\cite{longhi}. Examples of this include dynamic localization \cite{dynamic_localization}, Bloch oscillations\cite{bloch}, Zeno effect\cite{zeno} and Anderson localization\cite{anderson}. The main appeal of using optical systems is that experiments can be designed to focus on a particular aspect without the need to deal with the presence of many other effects commonly present in quantum solids, such as many-body effects. In optics it is also possible to steer and manage the propagation of excitations and to tailor the optical medium. Thus, it is no surprising that there have been also a number of recent works on BICs in classical optical systems\cite{optical_structures,yuri, molina}.

A different concept that has gained much recent attention is that of $\mathcal{P T}$-symmetry. It is based on the seminal work of Bender and coworkers\cite{bender1, bender2} who showed that non-hermitian Hamiltonians are capable of displaying a purely real eigenvalue spectrum provided the system is symmetric with respect to the combined operations of parity (P) and time-reversal (T) symmetry.  For one-dimensional systems the ${\mathcal{P T}}$ requirement leads to the condition that the imaginary part of the potential term in the Hamiltonian be an odd function, while the real part be even. In a $\mathcal{P T}$-symmetric system, the effects of loss and gain can balance each other and, as a result,  give rise to a bounded dynamics. The system thus described can experience a spontaneous symmetry breaking from a ${\mathcal{P T}}$ symmetric phase (all eigenvalues real) to a broken phase (at least two complex eigenvalues), as the gain/loss parameter is varied. In the case  of optics, 
the $\mathcal{P T}$-symmetry requirements lead to the condition that the real part of the refractive index be an even function, while the imaginary part be an odd function in space.
To date, numerous ${\mathcal{P T}}$-symmetric systems have been explored in several fields, from optics\cite{optics1,optics2,optics3,optics4,optics5}, electronic circuits\cite{electronic}, solid-state and atomic physics\cite{solid1,solid2}, to metamaterials\cite{engheta}, among others. The ${\mathcal{P T}}$ symmetry-breaking
phenomenon has been observed in several experiments\cite{optics5,experiments1,experiments2}.
\begin{figure}[t]
\centering
\includegraphics[width=0.45\textwidth]{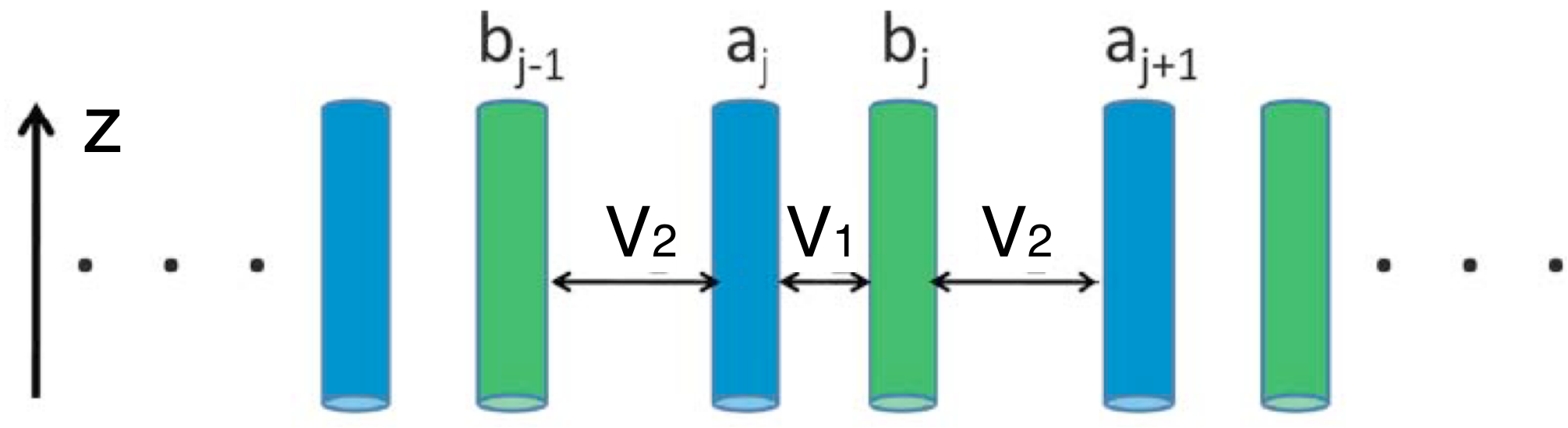}
\caption{Schematics of a binary waveguide
array with gain and loss. }
\label{fig1}
\end{figure}

In this paper, we merge together, for the first time to our knowledge, these two important concepts, namely
the concept of embedded states in the continuum and the concept of u symmetric systems in a single model.
We have considered a binary waveguide array with balanced gain and loss satisfying the PT symmetry conditions and
studied  the existence and properties of embedded states in the continuum for such systems.
We have demonstrated how to construct such bounded states and local potentials supporting them. We reveal that
for general envelope functions, the bounded states will force the system to undergo the PT symmetry
transition.

\section{Model}
Let us consider light propagating in a one-dimensional, binary waveguide array (Fig.\ref{fig1}), 
where the propagation constants at each waveguide are complex, with alternating signs for the imaginary parts.
The dynamical equations are 
\begin{eqnarray}
i {d a_{j}\over{d z}} + i \rho a_{j} + V_{1} b_{j} + V_{2} b_{j-1} & = & 0\nonumber\\
i {d b_{j}\over{d z}} - i \rho b_{j} + V_{1} a_{j} + V_{2} a_{j+1} & = & 0\label{eq:1}
\end{eqnarray}
where $j$ is the waveguide number, $a_{j}$ and $b_{j}$ are the mode  amplitudes at waveguides with gain and loss, respectively, $\rho$ is the rate of loss or gain, $V_{1,2}$ are the coupling coefficients and $z$ is the dimensionless propagation distance.

System (\ref{eq:1}) can be generated from the Hamiltonian
\begin{eqnarray}
H &=& -\sum_{j} [ i \rho (|a_{j}|^2 - |b_{j}|^2   ) + V_{1} a_{j}^{*} b_{j} + \nonumber\\
& & V_{2} a_{j}^{*} b_{j-1} + V_{1} b_{j}^{*} + V_{2} b_{j}^{*} a_{j+1} ]
\end{eqnarray}
through $i (d/d z) a_{j} = \partial H/\partial a_{j}^{*}$, $i (d/d z) b_{j} = \partial H/\partial b_{j}^{*}$, and is clearly non-hermitian due to the presence of the gain and loss coefficients.

\begin{figure}[t]
\centering
\includegraphics[width=0.225\textwidth]{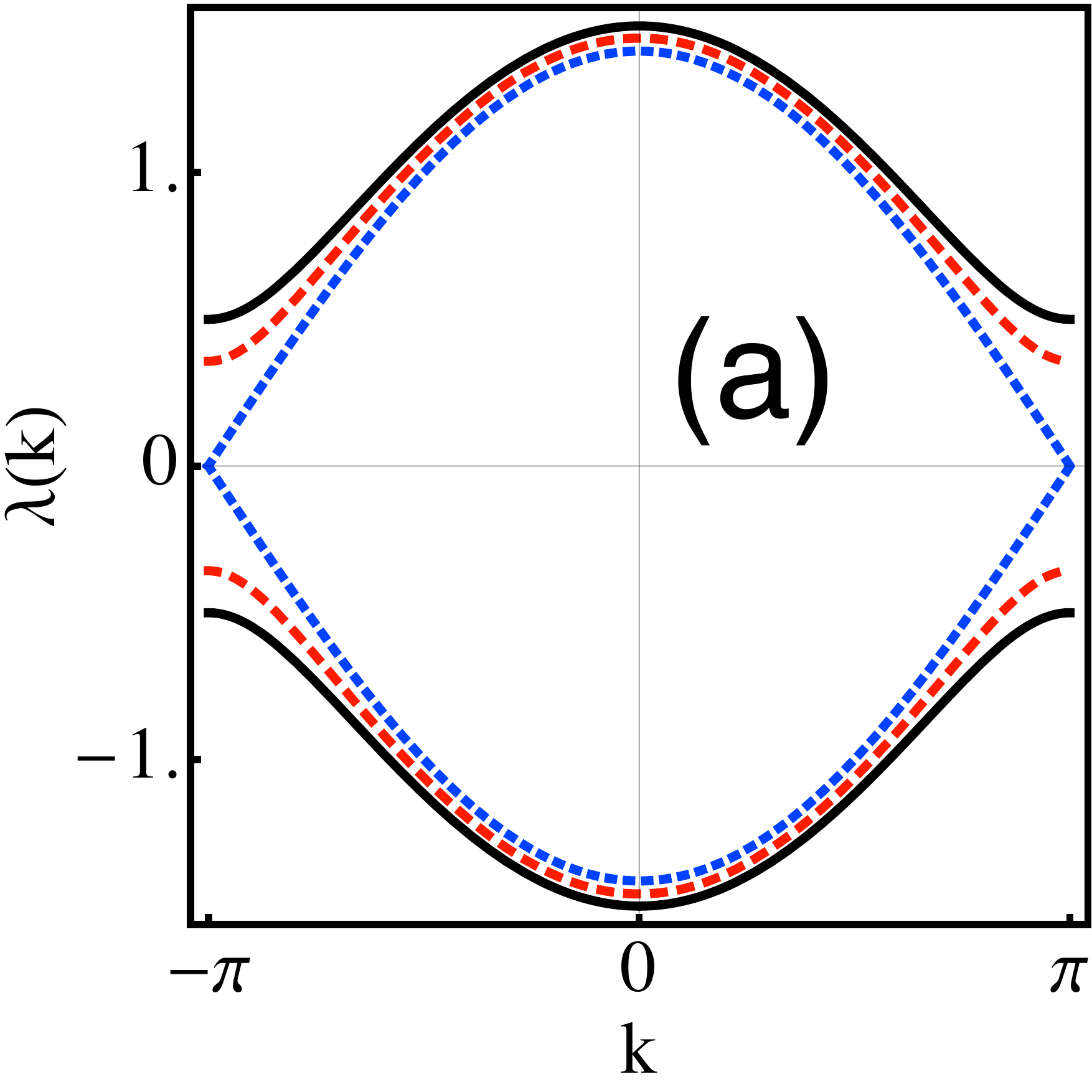}
\hspace{0.2cm}
\includegraphics[width=0.225\textwidth]{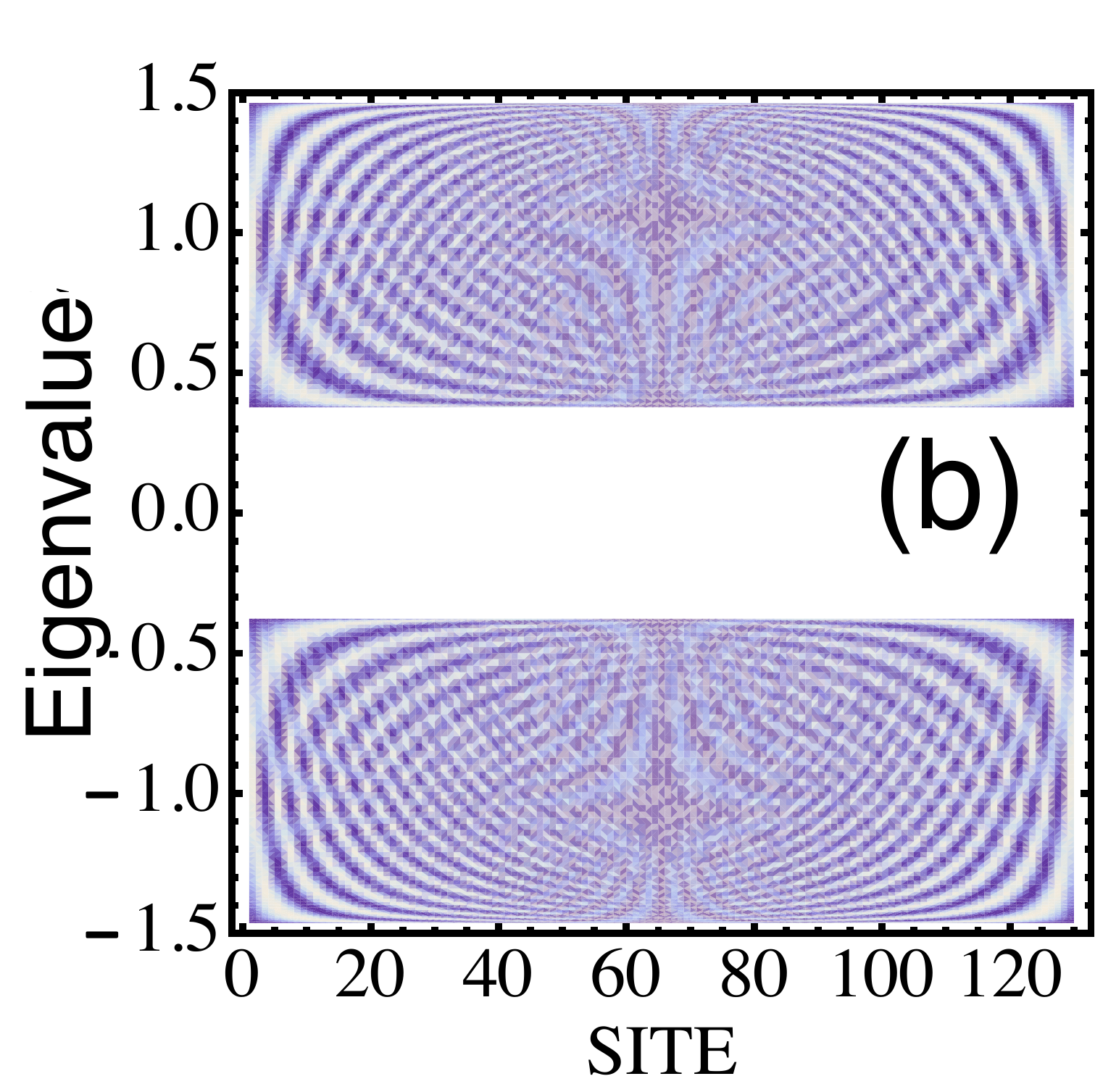}
\caption{(a) Dispersion relation for several gain and loss coefficients: $\rho=0$(solid), $0.25$(dashed) and $0.5$ (dotted). (b) Spatial distribution of the absolute value of complex eigenmodes ordered according to eigenvalue, for $\rho=0.33$. ($V_{1}=1, V_{2}=0.5$)}
\label{fig2}
\end{figure}

Let us look for the stationary modes of Eq.(\ref{eq:1}). Posing
Floquet-Bloch modes $(a_{j}, b_{j}) = (a_{0}, b_{0}) \exp(i \lambda z + i k j)$, one obtains
\be
\lambda(k) = \pm \left( V_{1}^2 + V_{2}^2 + 2 V_{1} V_{2} \cos(k) - \rho^2 \right)^{1/2}\label{eq:3}
\ee

which is real for all wavevector $k$ when
\be
|\rho| < |V_{1} - V_{2}|.
\ee
Figure \ref{fig2} shows how the dispersion relation changes
in the presence of the gain and loss coefficient. Most of the change happens at the band extremes, $k=\pm\pi$, where the bangap width decreases with $\rho$, vanishing at $\rho=|V_{1}-V_{2}|$. Figure \ref{fig2} also shows the spatial distribution of the eigenmodes in the presence of gain and loss. It is quite similar to the case of zero gain and loss.

At this point we should point out that the modes described by 
Eq.(\ref{eq:3}) are structurally unstable: If one changes the gain and loss coefficient so they are not equal to each other, but instead $\rho_{a}$ and $-\rho_{b}$, then it can be shown from Eq.(\ref{eq:1}) that all $\lambda$ become complex. More specifically, a new term $(i/2)(\rho_{a}-
\rho_{b})$ appears in front of the square root, while inside the root $\rho$ is replaced by $((\rho_{a}+\rho_{b})/2)^2$. Thus, modes with real propagation constants are only possible when $\rho_{a}=\rho_{b}=\rho$.

Let us now consider the problem of building a localized  eigenmode but with eigenenergy inside any of the allowed bands  of our binary lattice with gain and loss.  Following the prescription of  Wigner and von Neumann\cite{wigner} we select one of the eigenstates $\phi_{n}^{0}$, with eigenvalue $\lambda^{0}$, and proceed to modulate its envelope in such a way that the wave thus modulated is an eigenstate of the inhomogeneous system with eigenvalue $\lambda^{0}$:
\be 
C_{n}^{0} = \phi_{n}^{0} f_{n}
\ee
where $f_{n}$ is a decaying and normalizable envelope: $f_{n}\rightarrow 0$ as $n\rightarrow \infty$ and $\sum_{n} |f_{n}|^2 < \infty$. In order for this state to be an eigenstate of the system, we need to introduce a `local potential' $\epsilon_{n}$, whose shape will be adjusted to render $C_{n}^{0}$ as an eigenstate. This local potential is a site energy distribution that obeys the stationary equation
\begin {eqnarray}
(\epsilon_{n}-\lambda^{0} + i \rho) C_{n}^{0} + V_{1} C_{n+1}^{0} + V_{2} C_{n-1}^{0} & = & 0\hspace{0.5cm}\mbox{$n$ odd}\ \ \ \ \nonumber\\
(\epsilon_{n}-\lambda^{0} - i \rho) C_{n}^{0} + V_{2} C_{n+1}^{0} + V_{1} C_{n-1}^{0} & = & 0\hspace{0.5cm}\mbox{$n$ even}\ \ \ \ \label{eq:6}
\end{eqnarray}
where, without loss of generality, we have chosen the odd (even) sites as the one with gain (loss). Equation (\ref{eq:6}) can be formally solved for $\epsilon_{n}$:
\begin {eqnarray}
\epsilon_{n} &=& \lambda^{0} - i \rho - V_{1} (C_{n+1}^{0}/C_{n}^{0}) - V_{2} (C_{n-1}^{0}/C_{n}^{0})\hspace{0.5cm}\mbox{$n$ odd}\ \ \ \ \ \  \nonumber\\
\epsilon_{n} &=& \lambda^{0} + i \rho - V_{2} (C_{n+1}^{0}/C_{n}^{0}) - V_{1} (C_{n-1}^{0}/C_{n}^{0})\hspace{0.5cm}\mbox{$n$ even}\ \ \ \ \ \  
\label{eq:7}
\end{eqnarray}
where, $C_{n}^{0} = \phi_{n}^{0} f_{n}$. 

Let us first choose $f_{n}$ in the form $f_{n} = \sqrt{\phi_{n}^{0 *}/\phi_{n}^{0}}\ g_{n}$. With this choice, $C_{n}^{0}$ will be real if $g_{n}$ is real. We now take a monotonically decreasing envelope $g_{n}$ around some site $n_{0}$: 
\be 
\left( g_{n+1}\over{ g_{n}} \right) = (1 - \delta_{n})\hspace{0.5cm} (n>n_{0})
\ee
with $\delta_{n}<1$ to be determined later. From this, we can solve formally for $g_{n}$:
\be
g_{n} = \prod_{m=1}^{|n - n_{0}|-1} (1-\delta_{m})\label{eq:fn}
\ee
We can rewrite this as
\be g_{n} = \exp\left\{ \sum_{m=1}^{|n - n_{0}|-1}\log(1-\delta_{m})\right\}.
\ee
In the limit of large $n$, and using that $\delta_{m} < 1$, we can approximate $g_{n}$ as
\be
g_{\infty} \approx \exp\left\{ -\sum_{m=1}^{\infty}\delta_{m}\right\}.
\ee
Therefore, if we want $\lim_{\infty}g\rightarrow 0$, one must have $\sum_{m=1}^{\infty}\delta_{m}=\infty$.

Besides making sure to have a decreasing envelope, we must also make sure that the `local' decreasing potential remains bounded. From Eq.(\ref{eq:7}) we see that there could be divergences or near-divergences close to the zeroes of $C_{n}$. To avert that, we choose $\delta_{n}$ in the form
\be
\delta_{n} = {a\over{1 + |n|^b}} N^{2} |\phi_{n}^{0}|^2 |\phi_{n+1}^{0}|^2 \label{eq:12}
\ee
where $N$ is the number of lattice sites and $\phi_{n}$ is the (normalized) state being modulated, and $a$ and $b$ are adjustable parameters that will determine the rate of decay of the envelope and of the local potential that will support the embedded state. The shape of $\delta_{n}$ in Eq.(\ref{eq:12}) guarantees that at the zeroes of $\phi_{n}$, the local potential (Eq.(\ref{eq:7})) will be zero as well

With all of the above, the expression for the envelope function is
\be
f_{n} = \sqrt{{\phi_{n}^{0 *}\over{\phi_{n}^{0}}}}\  \prod_{m=1}^{|n - n_{0}|-1} \left\{ 1 - {a\over{1+|m|^b}} N^2 |\phi_{m}^{0}|^2 |\phi_{m+1}^{0}|^2 \right\}
\ee
and the envelope function is now complex, while the modulated state is real. Thus, we have built an eigenstate of the system that is localized in space but whose eigenvalue lies inside the continuous band, and the local potential that supports it.

\section{Localized states and their properties}

Figure 3 shows results for the case $N=132, V_{1}=1, V_{2}=1/2, \rho=0.33, a=0.4\ \mbox{and}\ b=0.9$. With these values the system is in the ${\mathcal{P T}}$-symetric regime since $\rho<\rho_{c}=|V_{1}-V_{2}| = 1/2$. 
Figures 3a(b) show the real(imaginary) part of the decaying envelope. Figure 3c shows the modulated state which oscillates in space but also decays away from $n_{0}=66$. Figure 3d shows the real of the local potential, respectively. The imaginary part of this potential can be read from Eq. (\ref{eq:7}) to be $(-1)^n \rho$.
\begin{figure}[t]
\begin{center}
\noindent
\includegraphics[scale=0.4]{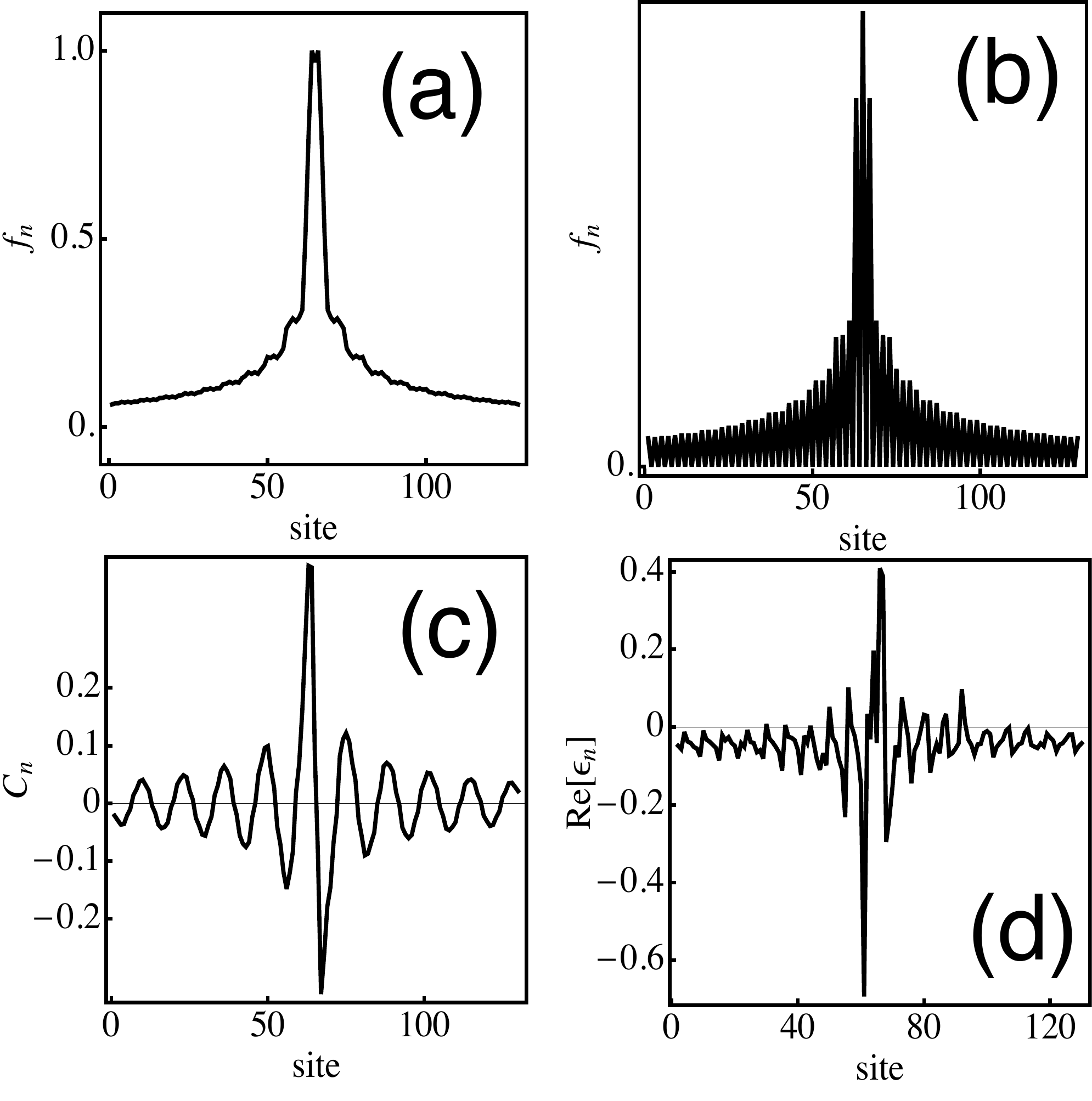}
\caption{(a) Real part of the envelope function. (b) Imaginary part of the envelope function.
(c) modulated state. (d) Real part of the site energy distribution.
($N=132, n_{0}=66, \lambda^{0}=1.31048, a=0.4, b=0.9$).}
\label{fig3}
\end{center}
\end{figure}

Let us consider now the effect of the modulation $f_{n}$ on the rest of the modes of the system. After the modulation on $\phi_{n}^{0}$ has been introduced, the rest of the system obeys the stationary equation
\begin{eqnarray}
(\epsilon_{n}-\lambda + i \rho) C_{n} + V_{1} C_{n+1} + V_{2} C_{n-1} & = & 0\hspace{0.5cm}\mbox{$n$ odd}\ \ \ \ \nonumber\\
(\epsilon_{n}-\lambda - i \rho) C_{n} + V_{2} C_{n+1} + V_{1} C_{n-1} & = & 0\hspace{0.5cm}\mbox{$n$ even}\label{eq:14}
\end{eqnarray}
where $\epsilon_{n}$ is given by Eq.(\ref{eq:7}). After inserting this into Eq.(\ref{eq:14}), one obtains
\begin{eqnarray}
\left\{ \lambda^{0} - \lambda - V_{1} {C_{n+1}^{0}\over{C_{n}^{0}}} - V_{2} {C_{n-1}^{0}\over{C_{n}^{0}}} \right\} C_{n} + V_{1} C_{n+1} + V_{2} C_{n-1}& = & 0\nonumber\\
\left\{ \lambda^{0} - \lambda - V_{2} {C_{n+1}^{0}\over{C_{n}^{0}}} - V_{1} {C_{n-1}^{0}\over{C_{n}^{0}}} \right\} C_{n} + V_{2} C_{n+1} + V_{1} C_{n-1}&=& 0.\ \ \ \ \ 
\end{eqnarray}
Rewriting this in a more compact form, we have that, after the modulation, the system obeys
\be
(-\lambda + \mu_{n}) C_{n} + V_{n,n+1} C_{n+1} + V_{n,n-1} C_{n-1} \label{eq:16}
\ee
where 
\be
\mu_{n} = \lambda^{0}- V_{n,n+1} \left({C_{n+1}^{0}\over{C_{n}^{0}}}\right) - V_{n,n-1} \left({C_{n-1}^{0}\over{C_{n}^{0}}}\right)\label{eq:17}
\ee
with $V_{n,n+1} = V_{n+1,n} = V_{1} (V_{2})$ for $n$ odd (even).
Since $\mu_{n}$ and $C_{n}^{0}$ are real, the hamiltonian of system (\ref{eq:16}) is hermitian, and all of its eigenvalues will be real. The modulation procedure has transformed a ${\mathcal{P T}}$-symmetric system from a non-hermitian one into a hermitian one. The gain and loss coefficients $\rho_{n}$ have now disappeared from view, although they are still contained in $\phi_{n}^{0}$ and $\lambda^{0}$. 

Figure 4a shows the absolute value of al eigenvectors of the system after the modulation, and we can clearly see the embedded mode at $\lambda = 1.31048$ which is localized in space, but surrounded by extended states. 
\begin{figure}[t]
\begin{center}
\noindent
\includegraphics[scale=0.26]{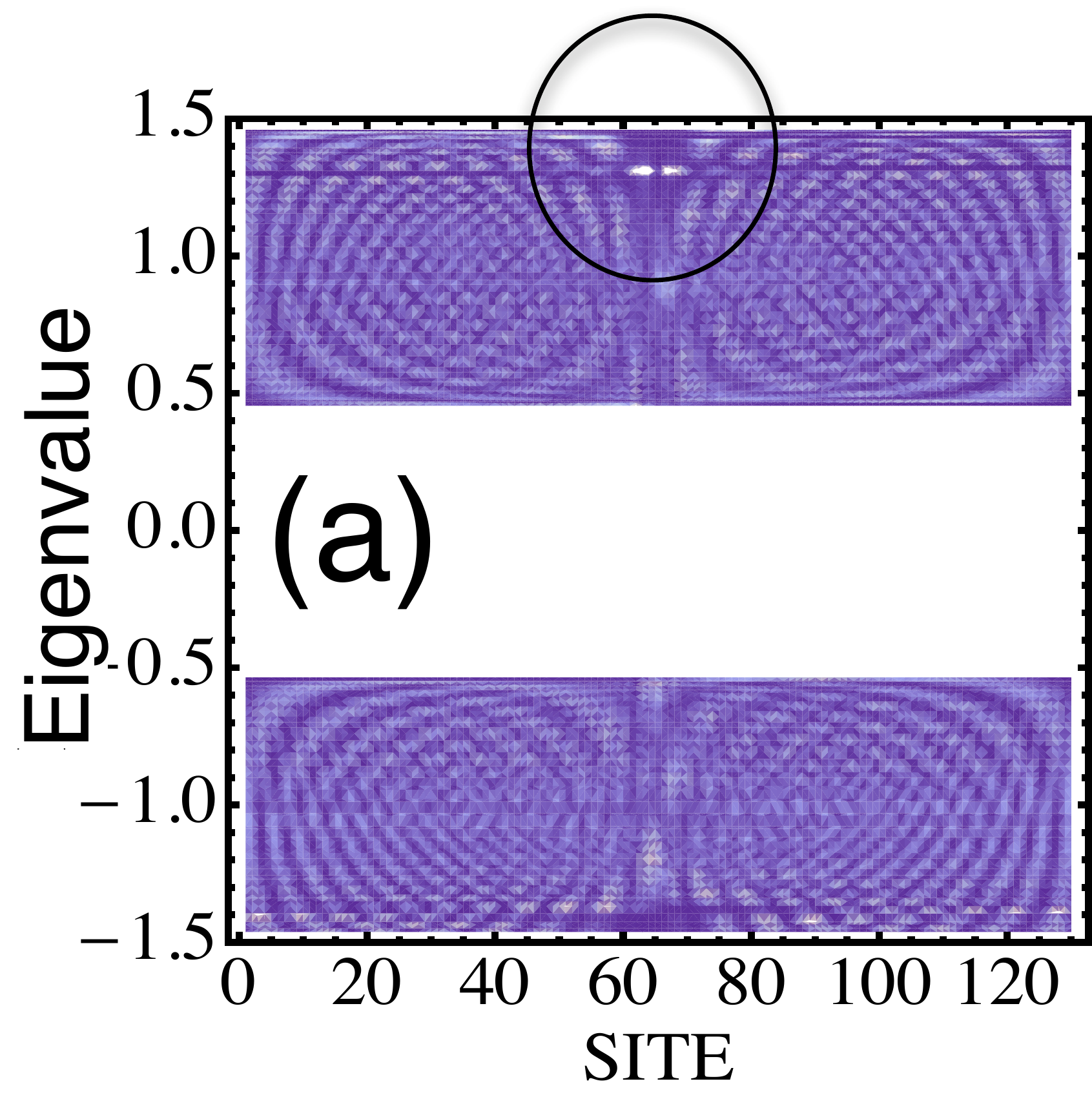}
\includegraphics[scale=0.32]{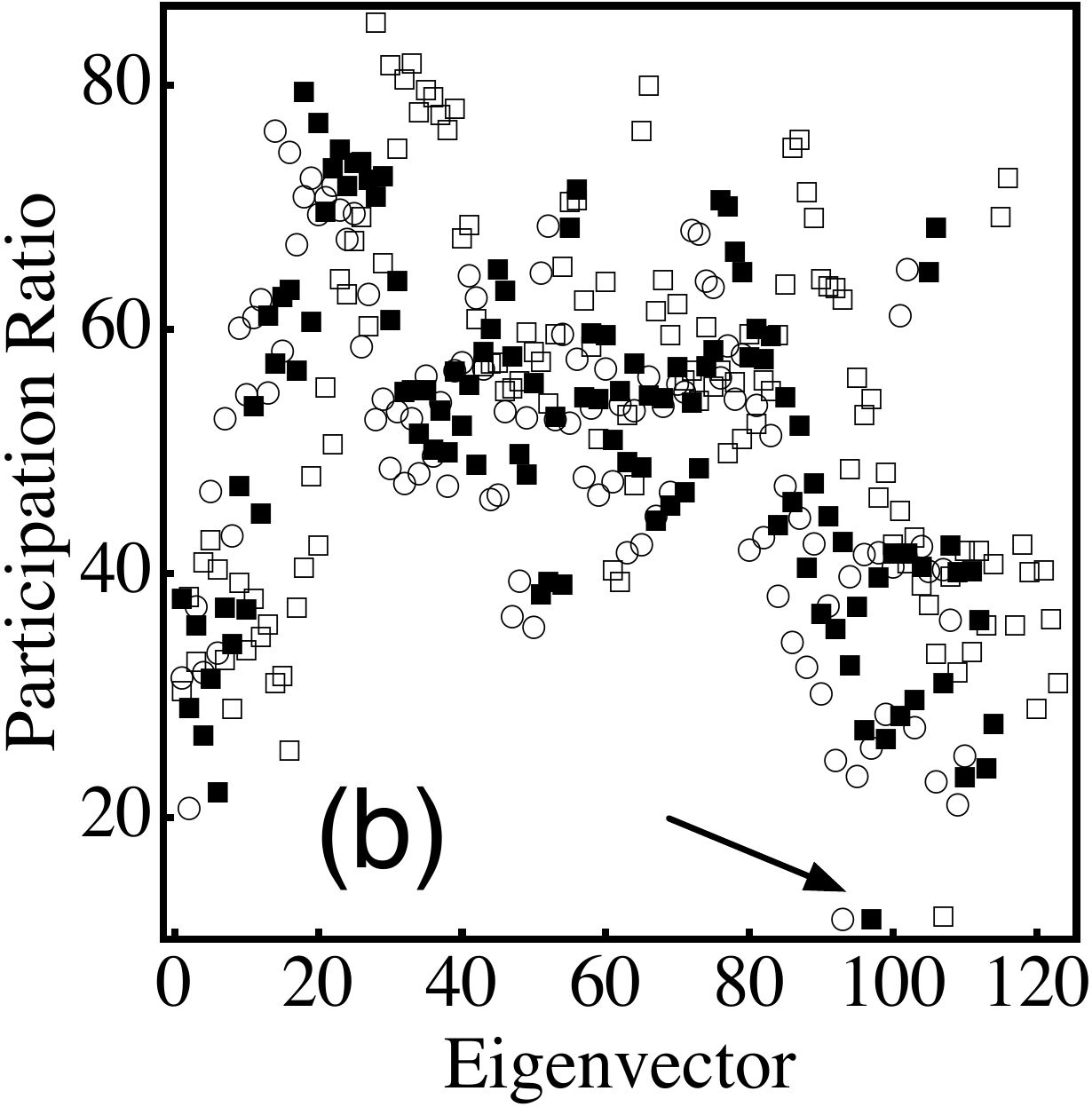}
\caption{(a) Absolute value of the eigenvectors of the modulated system in space, placed according to their eigenvalue. The circle indicates the position of the embedded mode. (b) Participation ratio of the eigenvectors of the  modulated system: For $\rho=0$ (clear squares),  $\rho=0.33$ (black squares), and $\rho=0.44$ (clear circles). The arrow indicates the value of $R$ corresponding to the embedded modes in the last two cases. 
($N=132, n_{0}=66, \lambda^{0}=1.31048\ \mbox{and}\ 1.27776, a=0.4, b=0.9$).}
\label{fig4}
\end{center}
\end{figure}
Not shown in the figure are a number of states that were pushed out of the band by the presence of the modulating potential, becoming impurity-like states. Now, a more rigorous measure of localization is provided by the participation ratio, $R$, defined by 
\be
R = (\sum_{n} |C_{n}|^2)^2/\sum_{n} |C_{n}|^4.
\ee

For a completely delocalized state, $R=N$, while for a completely localized one, $R=1$. Figure 4b shows $R$ for the system for increasing values of the gain and loss coefficient, inside the ${\mathcal{P T}}$-symmetric sector. In all cases we note a small decreasing tendency of $R$ as $\rho$ is increased. Even though the $R$ of the embedded states is the smallest, there are a umber of states with relatively small $R$, which indicates a tendency towards localization of the eigenstates of the modulated system. The states become `lumpier' but still extended.

Let us now choose a real envelope function $f_{n}$ instead of a real unmodulated state $C_{n}^{0}$. Following the procedure outlined before, we choose
\be
f_{n} =  \prod_{m=1}^{|n - n_{0}|-1} \left\{ 1 - {a\over{1+|m|^b}} N^2 |\phi_{m}^{0}|^2 |\phi_{m+1}^{0}|^2 \right\}
\ee
Figure 5 shows 
\begin{figure}[t]
\begin{center}
\noindent
\includegraphics[scale=0.4]{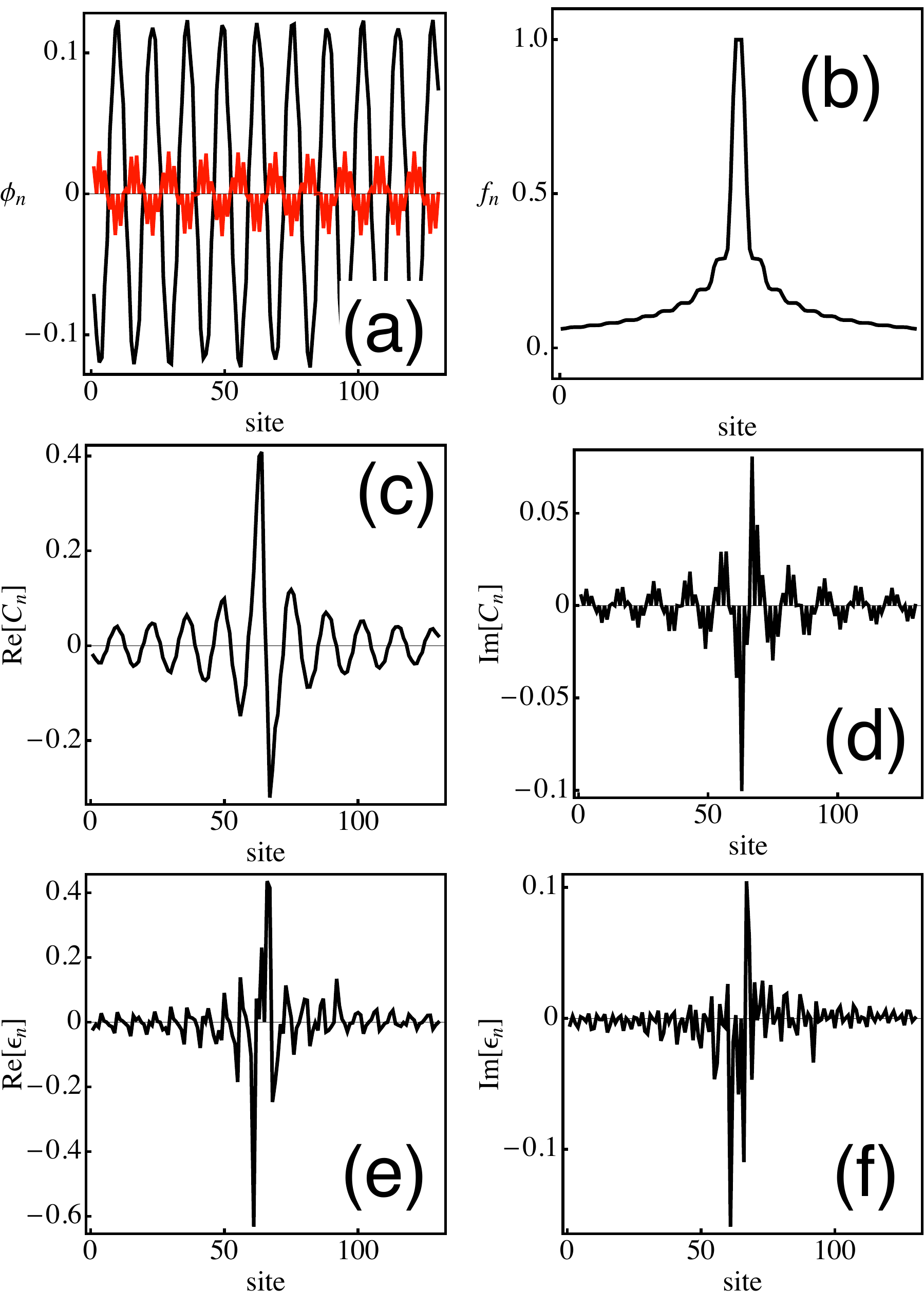}
\caption{(a) Unmodulated extended state. The large (small) amplitude oscillation corresponds to the real (imaginary) part. (b) Envelope function. (c) Real part
of the modulated state. (d) Imaginary part of the modulated state. (e) Real part of the site energy distribution. (f) Imaginary part of the site energy distribution.
($N=132, n_{0}=66, \rho=0.33, \lambda^{0}=1.31048, a=0.4, b=0.9$).}
\label{fig5}
\end{center}
\end{figure}
the (complex) unmodulated state, the envelope function, the real and imaginary parts of the modulated state, and the real and imaginary parts of the site energy distribution that supports the modulated state. Figure 6 shows the position of the modulated eigenstate inside of the (real part of the) band. We have left out those states that, after the modulation were pushed outside the band, becoming impurity states. The figure also shows the participation ratio of all states after the modulation. We notice that, as the gain and loss coefficient is increased, the participation ratio decreases, indicating a tendency towards greater localization, as in the previous case. Also contrary to the previous case,  there are a number of states whose participation ratio is smaller than the one corresponding to the embedded state. An examination of these modes reveals that most of them are also BICs that is, their eigenvalue lies inside the band, but are spatially localized. 

Now, unlike the previous case,  from Eqs.(\ref{eq:16}), (\ref{eq:17}), we have that in this case the final modulated system has a complex site energy distribution with a nontrivial imaginary part. This means that the system is now in a broken ${\mathcal{P T}}$ symmetry state.
\begin{figure}[t]
\begin{center}
\noindent
\includegraphics[scale=0.23]{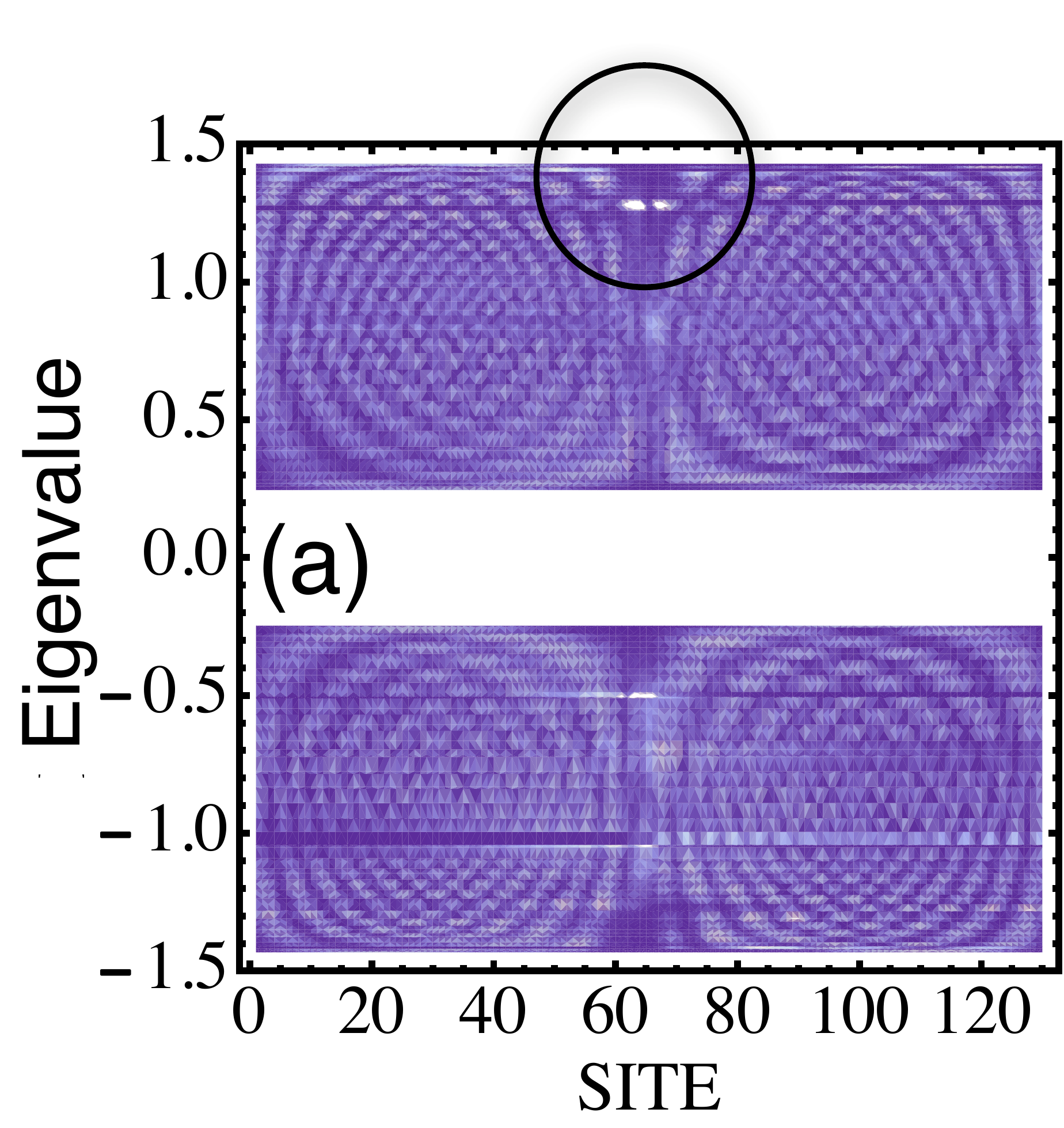}
\includegraphics[scale=0.24]{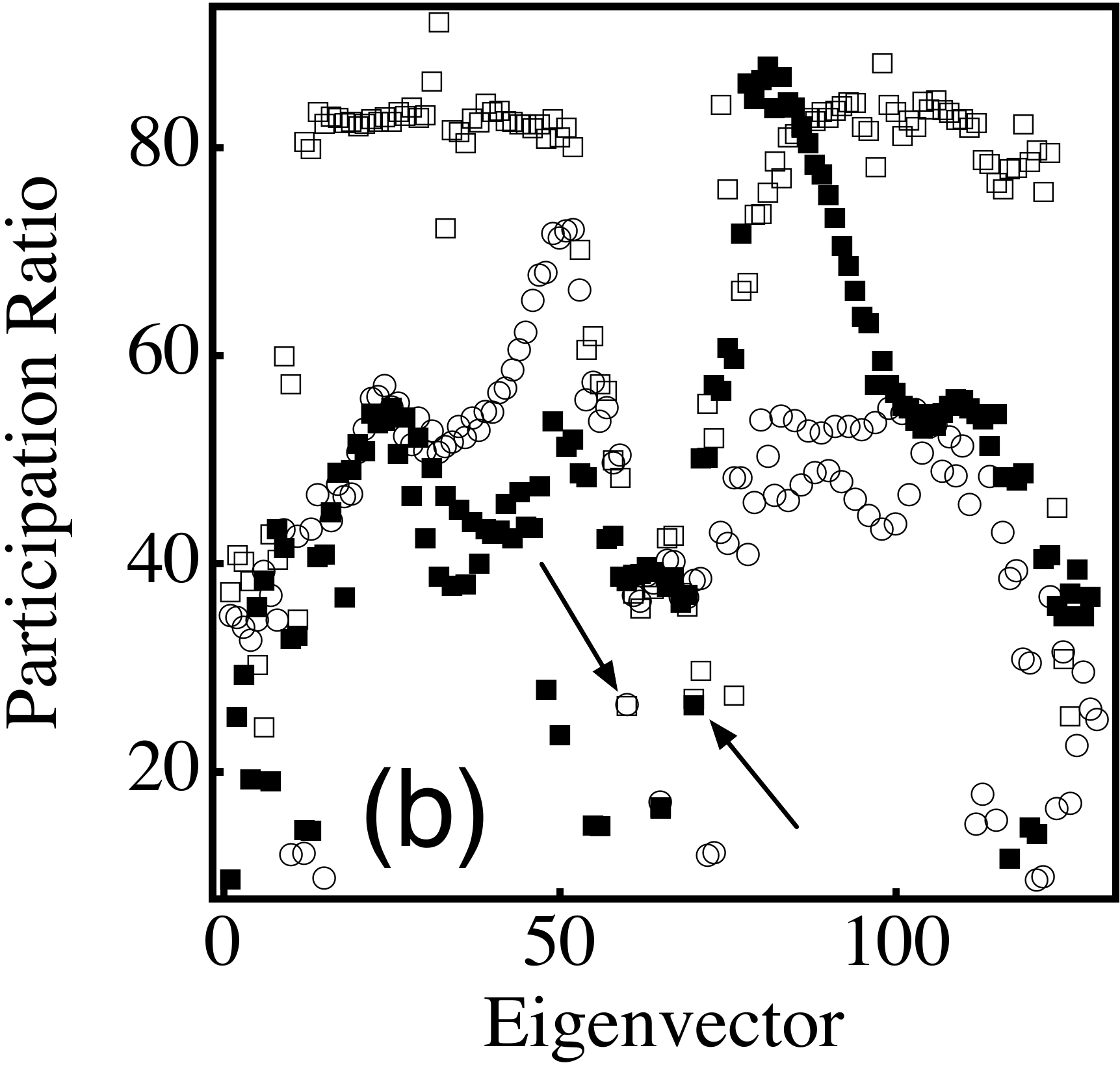}
\caption{(a) Absolute value of the eigenvectors of the modulated system in space, placed according to the value of the real part of their eigenvalues. The circle indicates the position of the embedded mode ($\rho=0.33, \lambda^0=0.4249$). (b) Participation ratio of the eigenvectors of the  modulated system: For $\rho=0,\lambda^0=-0.537958$ (clear squares),  $\rho=0.33,\lambda^{0}=0.4249$ (black squares), and $\rho=0.44, \lambda^0=-0.3095$ (clear circles). The left (right) arrow indicates the value of $R$ corresponding to the embedded mode with $\rho=0.33$ ($0.440$).($N=132, n_{0}=66, a=0.4, b=0.9$).}
\label{fig6}
\end{center}
\end{figure}
What happens if we start from our system already in the broken 
${\mathcal{P T}}$-symmetry state? This could happen, for instance, if the condition $\rho<|V_{1}-V_{2}|$ is not met, giving rise to some complex eigenvalues of the unmodulated system, for certain $k$-values. Or, the case when $\rho>V_{1}+V_{2}$, all of these eigenvalues are imaginary.
Figure 7 shows an 

example for one of the latest case, where we have use a real envelope function. As we can appreciate, there is no 
qualitative difference with the 
${\mathcal{P T}}$-symmetric case. The BIC formation seems very robust against changes in the ${\mathcal{P T}}$-symmetry regime.

\section{Conclusions}
We have analysed a possibility to combine two important recent concepts namely
the concept of bounded states in the continuum and the concept of ${\mathcal{P T}}$-
\begin{figure}[t]
\begin{center}
\noindent
\includegraphics[scale=0.58]{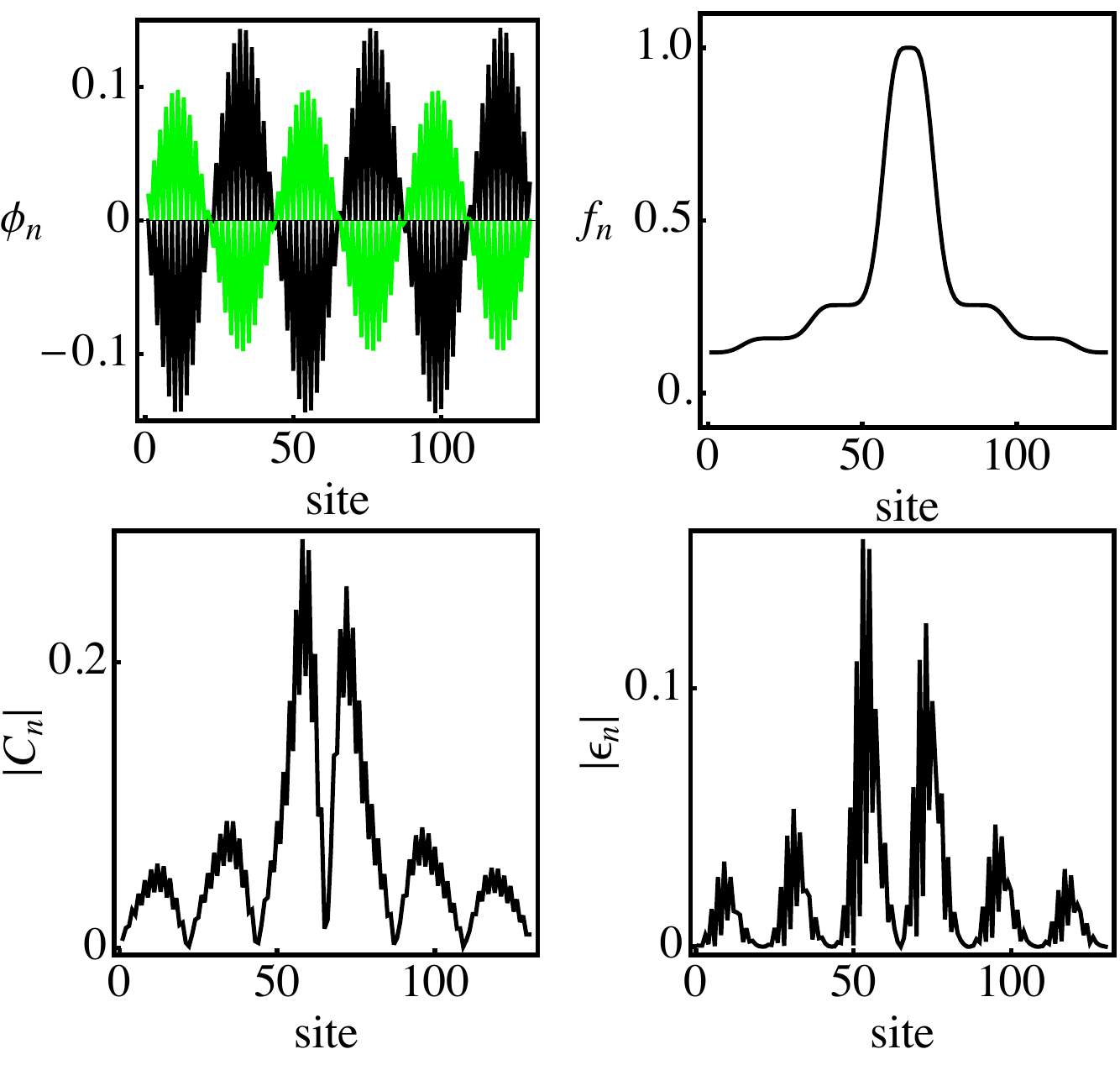}
\caption{Embedded mode in the ${\mathcal{P T}}$-broken symmetry regime. (a) Unmodulated extended state. The black (gray) amplitude oscillation corresponds to the real (imaginary) part. (b) Envelope function. (c) Absolute value 
of the modulated state. (d) Absolute value of the site energy distribution.
($N=132, n_{0}=66, \rho=1.6, \lambda^{0}=0.592381 i, a=0.4, b=0.9$).}
\label{fig7}
\end{center}
\end{figure}
symmetric systems. We
consider a binary waveguide array with balanced gain and loss satisfying the ${\mathcal{P T}}$ symmetry conditions and
pose a question about the existence of bounded modes in the continuum for such systems.
We have demonstrated how to construct such bounded states and local potentials supporting them.
We have revealed that  the process of creating a BIC will lead the system Ðoriginally in the
PT -symmetric regimeÐ to a Hermitian one if the modulated state is chosen as real; on the contrary if the modulated
state is chosen as complex, the system will go into the broken  ${\mathcal{P T}}$-symmetry regime after the creation of
the BIC. When the unmodulated state belongs to the broken ${\mathcal{P T}}$-symmetry phase, the BIC created is qualitatively
similar to their ${\mathcal{P T}}$-symmetric counterparts. Thus, for general envelope functions, the bounded states
will force the system to undergo the ${\mathcal{P T}}$ symmetry breaking transition.

This work was supported in part by Fondo Nacional de 
Ciencia y Tecnolog\'{\i}a (Grant 1120123), Programa Inicia- 
tiva Cient\'{\i}fica Milenio (Grant P10-030-F), Programa 
de Financiamiento Basal (Grant FB0824), and the Australian National University.

\end{document}